\documentstyle[12pt]{article}
\include{amssymb}
\begin{document}
\newcommand{\be}{\begin{equation}}
\newcommand{\ee}{\end{equation}}
\title{Why do we live in a Riemannian space-time ?}
\author{R. Scipioni, \footnote{scipioni@physics.ubc.ca}}
\maketitle
Department of Physics and Astronomy, The University of British Columbia,\\
6224 Agricultural Road, Vancouver, B.C., Canada V6T 1Z1
\begin{abstract}
\bigskip
We start from the pure Einstein-Hilbert action $S = \int \lambda^{2} R \star 1$ in Metric-Affine-Gravity, with the orthonormal metric $g_{ab} = \eta_{ab}$. We get an effective Levi-Civita  Dilaton gravity theory in which the Dilaton field is related to the scaling of the gravitational coupling.\\
When the Weyl symmetry is broken the resulting Einstein-Hilbert term is equivalent to the Levi-Civita one, using the projective invariance of the model, the non-metricity and torsion may be removed, so that we get a theory perfectly equivalent to General Relativity. This may explain why low energy gravity is described by a Riemannian connection.
\end{abstract}
\newpage
Among the four fundamental interactions, the two feeble are characterised by dimensional coupling constants, $G_{F} = (300 Gev)^{-2}$ and Newton's coupling constant $G_{N} = (10^{19} Gev)^{-2}$.\\
It is well known that interactions with dimensional coupling constants present many problems among which there is the renormalizability.\\
The success of the Weinberg-Salam model has told us that the weak interaction is characterised by a dimensionless coupling constant and the dimensions of $G_{F}$ are due to the spontaneous symmetry breaking mechanism, so that $G_{F} \cong \frac{1}{{v_{W}}^{2}}$ where $v_{W} \cong 300 Gev$ is the vacuum expectation value of the Higgs field.\\
The weakness of the weak interaction being related to the large vacuum expectation value of the scalar field [1].\\
It is believed that similar mechanisms may occur for gravity, which is characterised by a dimensionless coupling constant $\xi$. The weakness of gravity then would be related to the symmetry breaking at very high energies [2-4].\\
This may be obtained starting from a Dilaton theory which presents Weyl scale invariance. The potential $V(\psi)$ which appears in the action is assumed to have its minimum at $\psi = \sigma$, then when $\psi = \sigma$ the Dilaton theory reduces to the Einstein-Hilbert action with gravitational constant $G_{N} =\frac{1}{8 \pi \xi \sigma^2}$.\\
It has been shown that in the context of Metric-Affine-Gravity [5] the kinetic term for the dilaton may be obtained from a generalised Einstein-Hilbert term [6].\\
In this letter we continue the analysis of the model considerd in [6].\\ 
We investigate in the Tucker-Wang approach to non Riemannian gravity [6] the simple action:
\be
S = \int \lambda^2 R \star 1
\ee
Where $R$ is the scalar curvature associated with the full non Riemannian connection.\\
In the Tucker-Wang approach to MAG [7] we choose the metric to be orthonormal $g_{ab} = \eta_{ab} = (-1,1,1,1, ...)$ and we vary with respect to the coframe $e^{a}$ and the connection $\omega^{a}{}_{b}$ considered as independent gauge potentials.\\
We will study the two different cases where $\lambda$ is a dynamical variable subjected to aWeyl rescaling or the case when it is a constant.\\
We will prove that when the Weyl invariance is broken the theory obtained from (1) is perfectly equivalent to General Relativity. The breaking of Weyl symmetry may then give a giustification of why the low energy limit of gravity is Riemmanian.\\
 Before going into the calculations let us define how the Weyl rescaling transformations are defined in the Tucker-Wang approach.\\
Since the metric $g_{ab}$ is fixed we act only on the variable $\lambda$ and the coframe $e^{a}$ \footnote{We could have introduced a rescaling for the connection too, but in here we are looking for the simplest model so we consider the connection scale invariant}.\\
The Einstein-Hilbert term can be written in the form:
\be
R \star 1 = R^{a}{}_{b} \wedge \star (e_{a} \wedge e^{b})
\ee
Since the curvature two forms depends on the connection $\omega^{a}{}_{b}$, but not on the coframe it will not be affected by a rescaling of the coframe. The term $ \star (e_{a} \wedge e^{b})$ is a $n-2$ form. Then it is easy to see that if we introduce the rescaling defined by:
\begin{eqnarray}
\lambda \rightarrow \lambda \, \Omega^{p}\\ \nonumber
e^{a} \rightarrow e^{a} \, \Omega^{q}\\ \nonumber
\end{eqnarray}
The scale invariance of (1) holds if we satisfy the condition:
\be
2 \, p = - (n-2) \, q
\ee
In what follows we will suppose (4) to hold.\\
\\
If we consider the connection variation of (1) we get:
\be
D \star (e_{a} \wedge e^{b}) = - \frac{2}{\lambda}[d \lambda \wedge \star (e_{a} \wedge e^{b})] = A(\lambda)[d \lambda \wedge \star (e_{a} \wedge e^{b})]
\ee
with $A(\lambda) = -\frac{2}{\lambda}$.\\
\\
The full non Riemannian Einstein Hilbert term can be written as:
\begin{eqnarray}
R \star 1 = \stackrel{o}{R} \star 1 - {\hat{\lambda}}^{a}{}_{c} \wedge {\hat{\lambda}}^{c}{}_{b} \wedge \star (e^{b} \wedge e_{a}) - \\ \nonumber 
d( {\hat{\lambda}}^{a}{}_{b} \wedge \star (e^{b} \wedge e_{a}))
\end{eqnarray}
where ${\hat{\lambda}}^{a}{}_{b}$ is the traceless part of the non Riemannian part of the connection $\lambda^{a}{}_{b}$.\\

By considering the coframe variation we get then the generalized Einstein equations:\\
\begin{eqnarray}
\lambda^2 {\stackrel{o}{R}}^{a}{}_{b} \wedge \star (e_{a} \wedge e^{b} \wedge e_{c}) - 2 \lambda [ {\hat{\lambda}}^{a}{}_{b} \wedge d \lambda \wedge \star (e^{b} \wedge e_{a} \wedge e_{c})] \\ \nonumber
+ \lambda^2 [{\hat{\lambda}}^{a}{}_{d} \wedge {\hat{\lambda}}^{d}{}_{b}] \wedge \star (e_{a} \wedge e^{b} \wedge e_{c}) = 0 \\ \nonumber
\end{eqnarray}
The Cartan equation can be written as:
\be
D \star (e^{a} \wedge e_{b}) = A(\lambda)[d \lambda \wedge \star (e^{a} \wedge e_{b})] = F^{a}{}_{b}
\ee

We get from (8):
\be
f_{cab} = A(\psi) i_{c}( \star (d \lambda \wedge \star (e_{a} \wedge e_{b})))
\ee
the solution of which gives for the traceless part of the non-metricity and torsion:
\begin{eqnarray}
{\hat{Q}}^{ab} = 0 \\ \nonumber
{\hat{T_{c}}} = 0
\end{eqnarray}
and
\be
T = \frac{n-1}{2n} Q + \frac{1-n}{n-2} A(\lambda) d \lambda
\ee
the solution for the nonmetricity and torsion can then be written as:
\begin{eqnarray}
Q_{ab} = \frac{1}{n} g_{ab} Q \\ \nonumber
T^{a} = \frac{1}{2n}(e^{a} \wedge Q) - \frac{1}{n-2}(e^{a} \wedge d \lambda) A(\lambda)
\end{eqnarray}
Using the expression of $\lambda^{a}{}_{b}$ as a function of $T^{a}$ and $Q_{ab}$ [7]:
\be
2 \lambda_{ab} = i_{a}T_{b} - i_{b} T_{a} -(i_{a}i_{b} T_{c} + i_{b}Q_{ac} - i_{a} Q_{bc}) e^{c} - Q_{ab}
\ee
we obtain:
\be
\lambda_{ab} = -\frac{1}{2n}g_{ab} Q + \frac{1}{n-2} A(\lambda)(i_{a} (d \lambda) e_{b} - i_{b} (d \lambda) e_{a})
\ee
and the traceless part:
\be
{\hat{\lambda}}_{ab} = \frac{1}{n-2} A(\lambda)(i_{a} (d \lambda) e_{b} - i_{b} (d \lambda) e_{a})
\ee
By using the previous expression in the generalised Einstein equations we get after some calculations:\\
\bigskip
\be
\lambda^2 \, {\stackrel{o}{G}}_{c} - \beta [d \lambda \wedge i_{c} \star d \lambda + i_{c} d \lambda \wedge \star d \lambda] = 0
\ee
where ${\stackrel{o}{G}}_{c} = {\stackrel{o}{R}}^{a}{}_{b} \wedge \star (e_{a} \wedge e^{b} \wedge e_{c})$ and $ \beta = 4 \frac{n-1}{n-2}$ and the superscript (o) refers to the Levi-Civita part.\\
\\
The variation of (1) with respect to $\lambda$ gives:
\be
\lambda \, R \star 1 = 0
\ee
Which can be shown to be equivalent to:
\be
\lambda^2 \, \stackrel{o}{R} \star 1 - 4 \frac{n-1}{n-2} (d \lambda \wedge \star d \lambda) = 0
\ee
In conclusion we get the equations:
\begin{eqnarray}
 \lambda^2 {\stackrel{o}{G}}_{c}  - 4 \frac{n-1}{n-2}[d \lambda \wedge i_{c} \star d \lambda + i_{c} d \lambda \wedge \star d \lambda]= 0 \\ \nonumber
+  \lambda \stackrel{o}{R} \star 1 - \frac{4}{\lambda} \frac{n-1}{n-2}(d \lambda \wedge \star d \lambda) = 0 
\end{eqnarray}
For $n=4$ we have:
\begin{eqnarray}
\lambda^2 {\stackrel{o}{G}}_{c}  - 6 k [d \lambda \wedge i_{c} \star d \lambda + i_{c} d \lambda \wedge \star d \lambda] = 0 \\ \nonumber
+ \lambda \stackrel{o}{R} \star 1 - \frac{6}{\lambda} (d \lambda \wedge \star d \lambda) = 0 
\end{eqnarray}
The Einstein equations in (19) coincide with the conformally invariant Einstein equations obtained starting from the action:
\be
S = \int \lambda^2 \stackrel{o}{R} \star 1 + 4 \frac{n-1}{n-2} (d \lambda \wedge \star d \lambda) 
\ee
We have to remember however that this equivalence holds with the amendment that the Weyl rescaling is defined for the coframe and not for the metric since $g_{ab}$ is fixed to be orthonormal.\\
What has been said is valid because we are assuming that $\lambda$ may be affected by a Weyl rescaling.\\
Suppose now to consider the situation in which the Weyl symmetry is broken, we choose then a certain value of $\lambda$ :
\be
\lambda = \lambda_{0}
\ee
The Cartan equation then reduces to:
\be
D \star (e_{a} \wedge e^{b}) = 0
\ee
The solution of which is:
\begin{eqnarray}
Q_{ab} = \frac{1}{n} g_{ab}Q \\ \nonumber
T^{a} = \frac{1}{n-1}(e^{a} \wedge T) \\ \nonumber
T = \frac{n-1}{2n}Q \\ \nonumber
\lambda_{ab} = - \frac{1}{2n} g_{ab}Q \\ \nonumber
\end{eqnarray}
But what is more important is that:
\be
{\hat{\lambda}}_{ab} = 0
\ee
so that we get:
\begin{eqnarray}
R \star 1 \equiv \stackrel{o}{R} \star 1 \\ \nonumber
G_{c} = {\stackrel{o}{G}}_{c} \\ \nonumber
\end{eqnarray}
So the action and the field equations become equivalent to the Einstein theory obtained from the action:
\be
S = \int {\lambda_{o}}^{2} \stackrel{o}{R} \star 1
\ee

That is we get the vacuum Einstein equations:
\be
{\lambda_{o}}^{2} G_{c} = 0
\ee
with torsion and non-metricity given by (24).\\

In conclusion starting from the action $S = \int \lambda^{2} R \star 1$ in MAG we are able to exhibit two theories depending on whether $\lambda$ is a constant or a Weyl field variable. In the latter case we obtain a Dilaton-Levi-Civita model for the Einstein sector, in which $Q_{ab}, T^{a}, T$ are given by (11-12). In the former we get a vacuum Einstein theory with non-metricity and torsion (parametrised by $Q$) and related by (26). \\
The next step is to enquiry why $Q_{ab}$ and $T^{a}$ are zero in the broken phase.\\
To prove that we invoke the projective invariance [12,13] of action (1). Indeed action (1) is invariant under the projective transformation:
\be
\omega^{a}{}_{b} \rightarrow \omega^{a}{}_{b} + \delta^{a}{}_{b} \, P
\ee
with $P$ arbitrary 1-form.\\
Under this transformation $Q$ and $T$ transforms as:
\begin{eqnarray}
Q \rightarrow Q -2n \, P \\ \nonumber
T \rightarrow T + (1-n) \, P \\ \nonumber
\end{eqnarray}
If I choose $P = \frac{Q}{2n}$ then I get $Q' = 0$ and:
\be
T' = T + (1-n) \, P = T + \frac{1-n}{2n} Q
\ee
Which vanishes on account of the third of (24).\\
The conclusion is that the non-Riemmanian fields can be removed using a projective transformation and we are left with a theory completely equivalent to General Relativity.\\
The interesting issue of studying the stability against perturbation around $Q=T=0$ will be considered in another paper.\\
\\
I wish to thank the International center for cultural cooperation and development (NOOPOLIS) Italy for partial financial support, also the useful comments of R. Branderberger are acknowledged.\\
\newpage
\begin{center}
{\bf REFERENCES}
\end{center}
\bigskip
1] S. Weinberg, Phys. Rev. Lett. {\bf 19} 1264 (1967).\\
\\
2] A. Zee, Phys. Rev. Lett. {\bf 42}, 417 (1979).\\
\\
3] L. Smolin, Nucl. Phys. B {\bf 160} 223 (1979).\\
\\
4] S. L. Adler, Rev. Mod. Phys. {\bf 54}, 729 (1982).\\
\\
5] F. H . Hehl, J. D. McCrea, E. W. Mielke, Y. Ne'eman, Phys. Rep. {\bf 258} 1 (1995).\\
\\
6] R. Scipioni, Phys. Lett. A, {\bf 259} 2, 2 104 (1999).\\ 
\\
7]  R. Tucker, C. Wang, \emph{Non Riemannian Gravitational interactions}, Institute of Mathematics, Banach Center Publications, Vol. 41, Warzawa (1997), see also R. W. Tucker, C. Wang, Class. Quant. Grav. {\bf 12} 2587 (1995), and T. Dereli, R. W. Tucker, Class. Quant. Grav. {\bf 11} 2575 (1994).\\
\\
8] J. Kim, C. J. Park, and Y. Yonn, Phys. Rev. D {\bf 51} 562 (1995).\\
\\
9] J. Kim,, C. J. Park and Y. Yoon, Phys. Rev. D {\bf 51} 4595 (1995).\\
\\
10] Y. Yoon, Phys. Rev. D, {\bf D59} (1999) 127501.\\
\\
11] I. Benn, R. W. Tucker, \emph{An Introduction to Spinors and Geometry} (1987), Adam Hilger.\\
\\
12] J. P. Berthias, B. Shabid-Saless, Class. Quant. Grav. {\bf 10} 1039 (1993).\\
\\
13] A. Macias, E. W. Mielke, H. A. M. Tecoti, J. Math. Phys. {\bf 36} 10 (1995) 5868.\\
\\
\end{document}